# Magnetic-tunable nanoscale thermal radiation between twisted graphene gratings

Ming-Jian He[1,2], Hong Qi[1,2,*], Ya-Tao Ren[1,2], Yi-Jun Zhao[1], and Mauro Antezza[3,4,**]

1 School of Energy Science and Engineering, Harbin Institute of Technology, Harbin 150001, P. R. China

2 Key Laboratory of Aerospace Thermophysics, Ministry of Industry and Information Technology, Harbin 150001, P. R. China

3 Laboratoire Charles Coulomb (L2C), UMR 5221 CNRS-Université de Montpellier, F-34095 Montpellier, France

4 Institut Universitaire de France, 1 rue Descartes, F-75231 Paris, France

*Corresponding authors: Email: qihong@hit.edu.cn (H. Qi), mauro.antezza@umontpellier.fr (M. Antezza)

**Abstract:** This paper presents a comprehensive theoretical study of the magnetic-tunable near-field radiative heat transfer (NFRHT) between two twisted graphene gratings. As a result of the quantum Hall regime of magneto-optical graphene and the grating effect, three types of graphene surface plasmon polaritons (SPPs) modes are observed in the system: near-zero modes, high-frequency hyperbolic modes, and elliptic modes. The elliptic SPPs modes, which are caused by the combined effect of magnetic field and grating, are observed in the graphene grating system for the first time. In addition, the near-zero modes can be greatly enhanced by the combined effect grating and magnetic field, rendering graphene devices promising for thermal communication at ultra-low frequency. In particular, the near-zero modes result in a unique enhancement region of heat transfer, no matter for any twisted angle between gratings. The combined effect of grating and magnetic field is investigated simultaneously. By changing the strength of magnetic field, the positions and intensities of the modes can be modulated, and hence the NFRHT can be tuned accordingly, no matter for parallel or twisted graphene gratings. The magnetic field endows the grating action (graphene filling factors and twisted angles) with a higher modulation ability to modulate the NFRHT compared with zero-field. Moreover, the modulation ability of twist can be tuned by the magnetic field at different twisted angles. In sum, the combined effect of magnetic field and grating provides a tunable way to realize the energy modulation or multi-frequency thermal communications related to graphene devices.

# I. INTRODUCTION

Near-field radiative heat transfer (NFRHT), first proposed by Rytov [1] and formalized by Polder and Van Hove [2], can exceed Planck's blackbody limit by several orders of magnitude in nanoscale separation distance. In recent years, rapid progress has been achieved in nanoscale thermal radiation [3-14]. Graphene, a monolayer sheet of carbon atoms, exhibits intriguing electronic properties that arise from its massless Dirac dispersion of electrons. Owing to its unusual physical characteristics, graphene is a promising candidate for fundamental research and new applications [15-17].

Graphene surface plasmon polaritons (SPPs) can be excited in the near-field regime to transport energy in photonic channels [18]. In recent years, several theoretic studies on NFRHT concerning graphene have emerged. The near-field radiative heat transfer has been investigated between two suspended graphene sheets [19, 20], two graphene sheets patterned in ribbon arrays [21], graphene disks [22-25] and multilayer graphene systems [26-32]. Given its excellent modulating performance and strongly confined SPPs, graphene can greatly enhance and tune NFRHT when it is coated on dielectric materials [33-35] either in parallel-plate configurations [18, 26, 36-41] or grating geometry [42-48]. Moreover, graphene can enhance NFRHT for materials with nonmatching surface excitations [49]. Covering the cell of the near-field thermophotovoltaic system with a graphene monolayer enhances performance of the device [27, 36, 50-52]. However, existing research on modulating NFRHT with graphene is based on tuning the chemical potential of graphene by electronic method or doping.

In consideration of the magneto-optical characteristics of semiconductors, such as InSb, an external magnetic field is applied to modulate the NFRHT of magneto-optical nanoparticles [53-56] and plates [57, 58]. The unique phenomena caused by magnetic field are mainly attributed to the anisotropic permittivity of InSb. In addition, applying an external magnetic field perpendicularly to graphene induces some unique phenomena, such as quantum Faraday effects and Kerr effects [59, 60] because of the magneto-optical conductivities of graphene [61-63]. In particular, the magnetic field modulates the NFRHT between two suspended graphene sheets[64, 65]. Shubnikov–de Haas-like oscillations have been observed in the spectral heat flux caused by the intraband or interband transitions between various Landau levels (LLs). For a strong magnetic field, an additional band with the resonant peak appearing at near-zero frequency exists, which has never been found in graphene systems at such a low frequency before.

The surface characteristics of objects can be anisotropic when they are patterned into metasurfaces or gratings. In the near-field regime, the heat transfer can be modulated in a mechanical method by changing the twisted angle between the optical axes of the two anisotropic surfaces [66-70]. Considering this phenomenon, we combine the effects of twist and magnetic field in the present study to modulate the NFRHT between two graphene gratings. The paper is structured as follows. In Section II, the anisotropic magneto-optical conductivity of graphene gratings, reflection coefficients and energy transmission coefficients are introduced. In Section III, the NFRHT between two parallel gratings in the presence of a magnetic field is investigated. The effects of graphene filling factor and magnetic field on NFRHT are investigated. Then, in the absence of a magnetic field, the effect of twist on NFRHT is studied for different graphene filling factors. In Section IV, the combined effect of grating and magnetic field is studied simultaneously. Finally, in Section V, we provide conclusive remarks.

## II. PHYSICAL SYSTEM

### A. NEAR-FIELD RADIATIVE HEAT TRANSFER

Fig. 1

To explore the physical mechanism of heat transfer in graphene system, suspended graphene sheets [19, 28, 30], gratings [21, 47] and nanodisks [23, 71, 72] are considered as ideal systems to avoid the influence of other objects. The essential mechanism discovered in suspended graphene systems paves a way for improving the performance of realistic systems which are coated with graphene sheets or gratings. Moreover, the grating system owns another two manners to modulate the heat transfer, which are the orientation of the gratings and filling factors of graphene. Therefore, in the present work, we consider a system composed of two parallel graphene gratings separated by a vacuum gap with separation distance $d$. The two graphene gratings are composed of arrays of graphene strips with the periodicity $L$ and the strip widths $L_g$, as illustrated in Fig. 1. The temperatures of the two gratings are denoted as $T_1$ and $T_2$. The two gratings are twisted, and the twisted angle $\theta$ is represented by the main axes of the two gratings, which are labeled with blue and red dashed lines. An external magnetic field $B$ is applied perpendicularly to the graphene gratings. The net radiative heat flux between two graphene sheets is given by

$$\Phi = \int_0^\infty \frac{d\omega}{2\pi} \varphi(\omega) \tag{1}$$

where the monochromatic heat flux is

$$\varphi(\omega) = \frac{\hbar \omega n_{1,2}}{4\pi^2} \int_0^\infty \xi(\omega, \kappa) \kappa d\kappa \tag{2}$$

where $\hbar$ is Planck's constant divided by $2\pi$ and $n_{1,2}(\omega)=n_1(\omega)-n_2(\omega)$ denotes the difference between the two mean photon occupation numbers $n_{1/2}(\omega) = \left[\exp(\hbar\omega/k_B T_{1/2})-1\right]^{-1}$. $\kappa$ is the component of the wave vector parallel to the interface. $\xi(\omega,\kappa)$ is the energy transmission coefficient, which reads [67, 70, 73]

$$\xi(\omega,\kappa) = \begin{cases} \text{Tr}\left[\left(\mathbf{I}-\mathbf{R}_2^\dagger\mathbf{R}_2 - \mathbf{T}_2^\dagger\mathbf{T}_2\right)\mathbf{D}_{12}\left(\mathbf{I}-\mathbf{R}_1^\dagger\mathbf{R}_1 - \mathbf{T}_1^\dagger\mathbf{T}_1\right)\mathbf{D}_{12}^\dagger\right], & \kappa < \kappa_0 \\ \text{Tr}\left[\left(\mathbf{R}_2^\dagger-\mathbf{R}_2\right)\mathbf{D}_{12}\left(\mathbf{R}_1-\mathbf{R}_1^\dagger\right)\mathbf{D}_{12}^\dagger\right]e^{-2|\kappa_z|d}, & \kappa > \kappa_0 \end{cases} \quad (3)$$

where $\kappa_0=\omega/c$ is the wave vector in vacuum, $\kappa_z = \sqrt{\kappa_0^2 - \kappa^2}$ is the normal component of the wave vector in vacuum. $\mathbf{D}_{12} = \left(\mathbf{I}-\mathbf{R}_1\mathbf{R}_2 e^{2i\kappa_z d}\right)^{-1}$ is the Fabry-Perot-like denominator matrix and $\mathbf{R}_j$ ($j$=1, 2) is the reflection coefficient matrix for the $j$-th graphene grating, with the form

$$\mathbf{R}_j = \begin{bmatrix} r_j^{ss} & r_j^{sp} \\ r_j^{ps} & r_j^{pp} \end{bmatrix} \quad (4)$$

where the superscripts $s$ and $p$ represent the polarizations of transverse electric and transverse magnetic modes, respectively. The reflection coefficientss of graphene sheets coated on an isotropic substrate are given by [64, 65, 74]

$$\begin{aligned} r_{ss} &= \frac{(\rho_a+1)(1-\rho_c)+\rho_b\rho_d}{(\rho_a+1)(\rho_c+1)-\rho_b\rho_d} \\ r_{sp} &= \frac{2\rho_b}{(\rho_a+1)(\rho_c+1)-\rho_b\rho_d} \\ r_{ps} &= \frac{-2\rho_d}{(\rho_a+1)(\rho_c+1)-\rho_b\rho_d} \\ r_{pp} &= \frac{(\rho_a-1)(1+\rho_c)-\rho_b\rho_d}{(\rho_a+1)(\rho_c+1)-\rho_b\rho_d} \end{aligned} \quad (5)$$

where $\rho_a = \left(\sigma_{xx}/\omega\varepsilon_0 + \varepsilon_i/\kappa_{z,i}\right)\kappa_z$, $\rho_b = \sqrt{\mu_0/\varepsilon_0}\sigma_{xy}$, $\rho_c = \left(\sigma_{yy}\omega\mu_0 + \kappa_{z,i}\right)/\kappa_z$ and $\rho_d = \sqrt{\mu_0/\varepsilon_0}\sigma_{yx}$. $\varepsilon_0$ and $\mu_0$ are the permittivity and permeability of vacuum, respectively. $\kappa_{z,i} = \sqrt{\varepsilon_i\kappa_0^2 - \kappa^2}$ is the normal component of the wave vector in the substrate with the relative permittivity $\varepsilon_i$. The conductivities $\sigma_{xx}$, $\sigma_{yy}$, $\sigma_{xy}$, and $\sigma_{yx}$ are introduced in the following part.

**B. MAGNETO-OPTICAL CONDUCTIVITY OF GRAPHENE GRATINGS**

When a magnetic field is applied perpendicularly to graphene, the response of graphene electrons to the external magnetic field displays a characteristic optical quantum Hall effect [61, 75]. The conductivity of graphene becomes a tensor with nonzero values for diagonal and off-diagonal components, given by

$$\begin{pmatrix} \sigma_{xx} & \sigma_{xy} \\ \sigma_{yx} & \sigma_{yy} \end{pmatrix} = \begin{pmatrix} \sigma_L & \sigma_H \\ -\sigma_H & \sigma_L \end{pmatrix} \tag{6}$$

where $\sigma_L$ and $\sigma_H$ represent the longitudinal and Hall conductivities, respectively. Within the Dirac-cone approximation [76], the magneto-optical conductivity of graphene assumes the simple form in the random phase approximation [75, 77]

$$\sigma_{L(H)}(\omega, B) = g_s g_v \times \frac{e^2}{4h} \sum_{n \neq m = -\infty}^{+\infty} \frac{\Xi_{L(H)}^{nm}}{i\Delta_{nm}} \frac{n_F(E_n) - n_F(E_m)}{\hbar\omega + \Delta_{nm} + i\Gamma_{nm}(\omega)} \tag{7}$$

where $g_{s(v)}=2$ is the spin (valley) degeneracy factor of graphene, $e$ is the charge of an electron, and $h$ is Planck's constant. $n_F(E_n) = 1/\left[1 + e^{(E_n - \mu)/k_B T}\right]$ stands for the Fermi distribution function, $k_B$ is the Boltzmann constant, and $\mu$ and $T$ are the chemical potential and temperature of graphene, respectively. Throughout this paper, $T$ is fixed at 300 K unless otherwise noted. $\Gamma_{nm}(\omega)$ is the LL broadening, taken as 6.8meV in this paper [78].

With the application of an external magnetic field, electrons acquire considerable cyclotronic energies via the Lorentz force. The continuum Dirac energy spectrum evolves into degenerated LLs with the energies of $n$-th LL given by

$$E_n = \text{sign}(n)(\hbar v_F / l_B)\sqrt{2|n|} \tag{8}$$

where $v_F \approx 10^6$ m/s is the Fermi velocity of the carriers in graphene, $l_B = \sqrt{\hbar/eB}$ is the magnetic length, and $B$ is the intensity of the magnetic field. $\Delta_{nm} = E_n - E_m$ is the LL energy transition ($m, n = 0, \pm 1, \pm 2, \ldots$ are the LL indices). The matrix elements in Eq. (7) are given by

$$\begin{aligned} \Xi_L^{nm} &= \frac{\hbar^2 v_F^2}{l_B^2} \left(1 + \delta_{m,0} + \delta_{n,0}\right) \delta_{|n|-|m|,\pm 1} \\ \Xi_H^{nm} &= i\Xi_L^{nm} (\delta_{|m|,|n|-1} - \delta_{|m|-1,|n|}) \end{aligned} \tag{9}$$

where $\delta$ is the Kronecker symbol.

To demonstrate the quantum regime of the magneto-optical conductivity clearly, we show the schematic of the electronic transitions of graphene for $\mu$=0, 0.2, 0.3 eV with a magnetic field $B$=10 T ($E_1 \approx 0.11$ eV) in Figs. 2(a)-2(c), respectively. The figures show that the band structure of graphene is composed of two Dirac cones. The magnitudes of energy for different LLs can be regarded as the sizes of different plates in the Dirac cones. The number of last occupied LLs is denoted as $N_F$=int[$(\mu/E_1)^2$], which yields the number of occupied electron-degenerate LLs. In view of Eq. (8), $N_F$ undoubtedly increases as the chemical potential of graphene. The electronic transitions have two types: (i) intraband transitions limited to adjacent LLs ($N_F$ to $N_F+1$) within the positive cone; and (ii) interband transitions, connecting LLs from the negative cone with LLs in the positive cone. The allowed interband and intraband transitions are represented with the vertical arrows in blue and crooked arrows in red, respectively. As shown in Fig. 2(a), the state at $n=0$ falls at the apex of the Dirac cones, where positive and negative energy cones meet. The transition involving zero-energy from $n=0$ to $n=1$ LLs in Fig. 2(a) is the only transition that cannot be definitely designated as inter or intraband transition. In this study, it is denoted as an interband-like transition for simplicity.

In consideration that the magneto-optical conductivity of graphene is mainly determined by the longitudinal part, the real and imaginary parts of longitudinal conductivity are plotted for $B=10$ T, with red and blue solid curves in Figs. 2(d)-2(f). $\sigma_L$ is plotted as a function of $E/E_1$, with $E = \hbar\omega$ and $E_1 = \sqrt{2}\hbar v_F / l_B$. The three panels are for $\mu=0$, 0.2, 0.3 eV, corresponding to Figs. 2(a)-2(c). Intraband transition with an energy difference of $\Delta_{\text{intra}} = E_{N_F+1} - E_{N_F}$ dominates at low frequency where most of the spectral weight is centered around $\omega = \left(E_{N_F+1} - E_{N_F}\right)/\hbar = \left(\sqrt{N_F+1} - \sqrt{N_F}\right)E_1/\hbar$ (green dashed line in Fig. 2(e)). Interband transitions with an energy difference of $\Delta_{\text{inter}} = E_{n+1} + E_n \, (n \geq N_F)$ have a series of peaks at frequency $\omega = \left(E_{n+1} + E_n\right)/\hbar = \left(\sqrt{n+1} - \sqrt{n}\right)E_1/\hbar$ (blue dashed lines) beyond the interband threshold $E = 2\sqrt{N_F}E_1$ (violet dashed lines in Figs. 2(e) and (f)). A series of absorption peaks demonstrating the maxima of the real part and 0 for the imaginary part at high frequencies is observed.

In view of Eq. (8) and the definitive formula of $N_F$, $N_F$ decreases as the magnetic field $B$ increases. As a result, no intraband transitions can occur at a low electronic density or strong magnetic field (large $E_1$), i.e., $\mu=0$ eV$<E_1\approx 0.11$ eV. The magneto-optical conductivity is completely determined by the interband transitions. For a high electronic density of $\mu=0.2$ eV between $E_3$ and $E_4$ as shown in Fig. 2(e), some low-frequency interband transitions vanish and intraband transitions emerge. For a higher electronic density of $\mu=0.3$ eV as shown in Fig. 2(f), the interband threshold is so large that interband transitions decay greatly. In this case, intraband transitions dominate the magneto-optical conductivity of graphene. In sum, the response of graphene electrons to the external magnetic field has two regimes: (i) the semiclassical limit, for low magnetic field and/or high electronic density; and (ii) the quantum Hall regime, for strong magnetic field and/or low electronic density.

Fig. 2

For two twisted graphene gratings, the conductivity tensor should be modified by the effective conductivity [79]

$$\begin{pmatrix} \sigma_{xx} & \sigma_{xy} \\ \sigma_{yx} & \sigma_{yy} \end{pmatrix} = \begin{pmatrix} \sigma_{\|}\cos^2\phi_j + \sigma_{\perp}\sin^2\phi_j & \sigma_H^{eff} + (\sigma_{\|}-\sigma_{\perp})\sin 2\phi_j/2 \\ -\sigma_H^{eff} + (\sigma_{\|}-\sigma_{\perp})\sin 2\phi_j/2 & \sigma_{\|}\sin^2\phi_j + \sigma_{\perp}\cos^2\phi_j \end{pmatrix} \quad (10)$$

where the effective conductivities along ($\sigma_{/\!/}$) and across ($\sigma_\perp$) the main axis, and effective Hall conductivity $\sigma_H^{eff}$ are given by [79]

$$\sigma_\perp = \sigma_L \sigma_c / (f_g \sigma_c + f_c \sigma_L) \quad (11)$$

$$\sigma_H^{eff} = f_g \sigma_\perp \sigma_H / \sigma_L \quad (12)$$

$$\sigma_{\|} = f_g \sigma_L + f_g \sigma_H^2 / \sigma_L - \left(\sigma_H^{eff}\right)^2 / \sigma_\perp \quad (13)$$

where $f_g = L_g/L$ is the graphene filling factor, and $f_c = 1 - f_g$ is the filling factor of the free space between adjacent strips. $\sigma_c = -2i\omega\varepsilon_0\varepsilon_{eff}\frac{L}{\pi}\ln\left[\csc\left(\frac{\pi f_c}{2}\right)\right]$ [80] is an equivalent conductivity associated with the near-field coupling between adjacent strips, where $\varepsilon_{eff}$ denotes the relative permittivity of the dielectric medium surrounding graphene. For two suspended graphene sheets or gratings, $\varepsilon_{eff}$ equals 1.

For the strip periodicity $L$=20 nm and graphene filling factor $f_g$=0.5, the imaginary parts of anisotropic optical conductivities for orthogonal directions ($\sigma_{/\!/}$ and $\sigma_\perp$) are plotted with blue and red lines in Figs. 2(g)-2(i) for $\mu$=0, 0.2, and 0.3 eV, respectively. Only when Im[$\sigma_{/\!/}$] is positive can graphene SPPs work. For low chemical potentials ($\mu$=0 eV), the magneto-optical conductivity of graphene gratings also reveals the multiband spectrum. Moreover, they match well with the frequencies defined by the interband transitions, represented by blue dashed lines. No essential changes are observed in the magneto-optical conductivity of graphene gratings, except that they convert to anisotropy. However, for larger chemical potentials of $\mu$=0.2 eV and 0.3 eV when intraband transitions dominate, below the interband threshold, the effective conductivities are similar to those of graphene gratings in the absence of a magnetic field [21]. The oscillations of Im[$\sigma_{/\!/}$] at high frequencies above the interband threshold imply that the interband transitions are not greatly affected by grating. The phenomenon implies that the magnetic field exerts minimal effect on conductivities when graphene is patterned into gratings for high electronic density. In other words, for large chemical potentials, the pattern transformation from sheets to gratings renders the magneto-optical conductivity of graphene impossible to exhibit. Therefore, investigating the NFRHT for graphene gratings with high chemical potentials is unnecessary.

## III. RESULTS AND DISCUSSION

To ensure the accuracy of the calculations based on the effective medium theory (EMT), the vacuum gap distance $d$ should be several times larger than the strip periodicity. As given by Ref [21], for $L$=20 nm, the EMT predicts the real heat flux well when $d \geq 60$ nm. In the following results, the parameters are selected as $d$=60 nm, $L$=20 nm, $T$=300 K unless otherwise noted. $\phi_j$ in Eq. (10) denotes the azimuthal angle of the main axis for grating $j$ ($j$=1, 2), which is illustrated in Fig. 1. When the gratings are aligned, the relationship of azimuthal angles for the two gratings is $\phi_2$= -$\phi_1$. If a twisted angle $\theta$ exists between the two gratings, then $\phi_2$=- $\phi_1$+ $\theta$.[73]. In view of Eqs. (4)–(5) and (10), integration over azimuthal angles is necessary for the energy transmission coefficients because Fresnel's coefficients and conductivity of graphene gratings depend on $\phi$. Hence, the energy transmission coefficients considering the integration over azimuthal angles are given as

$$\xi(\omega,\kappa) = \frac{1}{2\pi}\int_0^{2\pi} \xi(\omega,\kappa,\phi)d\phi \tag{14}$$

In particular, when $f_g$=1, our results match well with those for two magnetic graphene sheets[64, 65]. When $f_g$<1, $B$=0 T and twisted angle $\theta$=0°, we can obtain the same results as those for graphene gratings in the absence of a magnetic field [21]. In the following discussion, the radiative heat transfer coefficient (RHTC) is utilized to evaluate the NFRHT between two twisted graphene gratings as

$$h = \frac{\Delta \Phi}{\Delta T} = \frac{1}{8\pi^3}\int_0^\infty \hbar\omega \frac{\partial n}{\partial T} d\omega \int_0^{2\pi}\int_0^\infty \xi(\omega,\kappa,\phi)\kappa d\kappa d\phi \tag{15}$$

### A. PARALLEL GRATINGS IN MAGNETIC FIELD

Fig. 3

The effect of magnetic field on heat transfer between two parallel gratings is studied. The RHTCs corresponding to different sets of ($B,\mu$) are given in Fig. 3(a) for parallel gratings with graphene filling factor $f_g$=0.5. The results reveal notable differences for different chemical potentials of graphene. The maximum RHTC is reached around $\mu$=0.15–0.2 eV, and its trend varies with magnetic field at different $\mu$. The modulation factor of magnetic fields is defined as $\eta(B)=(h(B,\mu)-h(0,\mu))/h(0,\mu)$ to demonstrate the modulation effect of magnetic field on heat transfer quantitatively. The results of $\eta(B)$ for different chemical potentials are given in Fig. 3(b). The magnitude of $\eta(B)$ can reflect the changes in RHTC in the presence of a magnetic field, which is viewed as the modulation ability of the magnetic field. The gradients of the $\eta(B)$ curves at different $B$ values are regarded as the modulation speed for the specific magnetic field. The first glimpse of the $\eta(B)$ results reveals the general trend that the modulation ability of the magnetic field is weakened as the chemical potentials increase. For $\mu$=0–0.04 eV, the heat transfer is greatly weakened due to the presence of a magnetic field. $\eta(B)$ monotonically decreases as the magnetic field when $\mu$ is 0 eV and 0.01 eV. Moreover, the modulation speed of the magnetic field decays as $B$ increases. $\eta(B)$ has a much lower magnitude at $\mu$=0.1, 0.2, 0.3 eV than at lower chemical potentials. The magnetic field has almost no effect on RHTC when $\mu$=0.3 eV. This result is mainly attributed to the failure of the magnetic field to lead to the magneto-optical conductivity for graphene gratings, as observed in Fig. 2. The above results indicate that the modulation effect of magnetic field has different levels and trends corresponding to different $\mu$. The heat transfer can be greatly influenced by the magnetic field when the chemical potentials of graphene are low, especially for $B$=0–5 T. For this reason, our research results are mainly focused on the low chemical potentials of graphene $\mu$=0.01 eV.

Fig. 4

The contours of energy transmission coefficients for different $B$ are given in Fig. 4(a) to explain the modulation effect of magnetic field. For zero-field in Fig. 4(a-1), the energy transmission coefficients occupy a whole $\omega$-$\kappa$ area, which is similar to that of two suspended graphene sheets. Then, they are converted to two separated parts when a magnetic field is applied. In fact, energy transmission coefficients split into multiband spectra in higher frequency. However, they are so weak that we do not demonstrate them here. As $B$ increases, the low-frequency modes gradually decay to a narrow band, and they are given in separate panels in Figs. 4(a-4)-(a-6). To demonstrate the trend of high-frequency and low-frequency modes clearly, the spectral RHTCs for $B$=5, 10, 15 T are given in Fig. 4(b). When $B$ is higher than 5 T, the low-frequency modes become near-zero modes, as demonstrated in magnetic graphene sheets [65]. The near-zero modes gradually become stronger and the band becomes wider as *B* increases. Instead, the high-frequency modes gradually attenuate and show a blue shift at the same time. The two completely different trends of near-zero modes and high-frequency modes result in the slowdown of modulation speed for magnetic field, which is observed in Fig. 3(b) for strong magnetic field. Fig. 4(c) shows a plot of the imaginary part of effective conductivities along the main axis Im[$\sigma_{//}$] to explain the existence of near-zero modes and the blue shift of high-frequency modes. Im[$\sigma_\perp$] remains negative in the whole frequency regime of interest for different *B*; thus, it is not given in the plot. Only when Im[$\sigma_{//}$] and Im[$\sigma_\perp$] are positive and negative, respectively, can hyperbolic graphene SPPs work. As shown in Fig. 4(c), the zero points for *B*=5, 10, 15 T are marked with dashed lines together with arrows and are added to the energy transmission coefficient contour. The blue shift can be explained by the trend of Im[$\sigma_{//}$] with respect to increasing *B*. The inset in Fig. 4(c) denotes the local enlarged drawing of Im[$\sigma_{//}$] near-zero frequency. It demonstrates that Im[$\sigma_{//}$] is positive in a narrow band near zero-frequency for larger *B*. The unique trend of near-zero modes observed in Fig. 4(b) can be well explained by the positive Im[$\sigma_{//}$] at low frequency.

Fig. 5

For two parallel gratings, the graphene filling factor greatly influences heat transfer. To study the effect of the graphene filling factor $f_g$ together with the magnetic field *B*, we provide the RHTCs corresponding to different sets of (*B*, $f_g$) in Fig. 5(a). For different $f_g$, the RHTC decreases with increasing *B* and the modulation speed of magnetic field is larger for the weak fields, which is observed in Fig. 3(b). In the presence of a weak magnetic field, the maximum RHTC is reached when $f_g$=1, which means graphene sheets. To demonstrate the effect of $f_g$ clearly, we plot $\eta$ ($f_g$)=(*h*(*B*, $f_g$)-*h*(*B*,1))/*h*(*B*,1), defined as the the modulation factor of graphene filling factor, as a function of $f_g$ for different *B* in Fig. 5(b). The results show clearly that the magnetic field can tune the effect of graphene filling factor on modulating NFRHT. For zero-field, $\eta$ ($f_g$) ranges from -0.7 to 0, but it has a wider range when a magnetic field is applied. The modulation range increases by nearly 30% when *B* increases from 0 to 5 T. The larger modulation range results from a higher modulation ability for heat transfer when the magnetic field and grating effects both work. As shown in Fig. 5(b-1), in the presence of a weak magnetic field, the modulation effect of $f_g$ is greater for larger *B* at low $f_g$ (0–0.9). Nevertheless, for large $f_g$ (0.9–1), the $\eta$ ($f_g$) curves for different *B* are nearly in coincidence. It means that the grating effect dominates the heat transfer and that the effect of magnetic field loses efficacy when graphene strips greatly occupy the grating configuration. In Fig. 5(b-2), for strong magnetic fields $B$=5–15 T, the $\eta(f_g)$ curves have different trends with growing $B$, when the effect of the magnetic field is greater than that of the weak magnetic field. As $B$ increases, the modulation speed of $f_g$ increases at low $f_g$ (0.1–0.6) and is nearly unchanged at large $f_g$ (0.6–0.8). Interestingly, $\eta(f_g)$ non-monotonically increases with $f_g$, especially in the presence of a strong magnetic field (10–15 T) at large $f_g$ (0.8–1). The unexpected enhancement caused by grating action is observed in the magnetic graphene system for the first time. The inset in Fig. 5(b-2) shows the detailed view of the unique phenomenon. The maximum enhancement exists at around $f_g$=0.9 for $B$=15 T and the enhancement gradually vanishes when $B$ is lower than 10 T. The unique phenomenon is caused by the combined effect of magnetic field and grating, and the physical mechanism is discussed in the following part in detail.

Fig. 6

The spectral results for different $f_g$ with $B$ =15 T are given in Fig. 6(a) to explore the physical mechanism of the grating effect. When graphene sheets are patterned into strips, the near-zero peak greatly increases and the band becomes wider as $f_g$ decreases. Moreover, the high-frequency peak is converted into two parts, and they have different trends with $f_g$. To intuitively explain the two modes, we plot the imaginary parts of anisotropic optical conductivities for orthogonal directions in Fig. 6(b). Only when Im[$\sigma_{//}$] is positive can graphene SPPs work, which is denoted by the red vertical line. Hyperbolic and elliptic modes come into play when Im[$\sigma_\perp$] is negative and positive, respectively. The bands of hyperbolic and elliptic modes are labeled with vertical lines in Fig. 6(b). On the right side of the red vertical line, the dividing point between hyperbolic and elliptic modes are denoted with the blue and violet vertical lines for $f_g$=0.9 and 0.8, respectively. As $f_g$ decreases, the position of the dividing point shows a blue shift and Im[$\sigma_\perp$] for the elliptic modes is very close to zero when $f_g$=0.6. According to the above phenomenon, the two parts observed in Fig. 6(a) are hyperbolic and elliptic graphene SPPs modes. The contours of energy transmission coefficients for different $f_g$ are given in Fig. 6(c) to verify the two modes. The contours match with the change trend of spectral RHTCs. Using the relationships $\xi(\omega,\kappa,\phi)=\xi(\omega,\kappa_x,\kappa_y)$, $\kappa_x = \kappa\cos\phi$, and $\kappa_y = \kappa\sin\phi$, we provide the anisotropic energy transmission coefficients for the specific frequencies in panels (i)–(iii). The dispersion relations are represented with red dashed lines, which are calculated by zeroing the denominator of Eq. (5). The frequencies of panels (i)–(iii) are indicated with dashed lines in Figs. 6(c-1) and (c-2). For $f_g=1$, which means graphene sheets (Im[$\sigma_\perp$] and Im[$\sigma_{//}$] are in coincidence in Fig. 6(b)), the energy transmission coefficients are isotropic in panel (i). Panels (ii) and (iii) verify that the low-frequency and high-frequency modes are hyperbolic and elliptic graphene SPPs modes, respectively. Elliptic SPPs modes are observed in the graphene grating system for the first time, which is caused by magnetic field and grating simultaneously. Moreover, according to Figs. 6(a) and 6(c), as $f_g$ decreases, the hyperbolic modes gradually enlarge and the peak position is nearly unchanged, whereas the elliptic modes dramatically weaken and show a blue shift. When $f_g=0.6$, the elliptic modes almost vanish. The attenuation and blue shift of the elliptic modes are attributed to the positive–negative transition and the zero point of Im[$\sigma_{//}$] observed in Fig. 6(b), respectively. The unexpected enhancement observed in Fig. 5(b) can be well explained by the great enhancement of near-zero modes. When $f_g$ is lower than 0.6, the near-zero modes and high-frequency hyperbolic modes both decay as $f_g$ decreases, which results in the monotonic trend of $\eta$ ($f_g$) in Fig. 5(b).

## B. TWISTED GRATINGS WITHOUT MAGNETIC FIELD

Fig. 7

In this section, the effect of graphene filling factors and twisted angles between two graphene gratings on heat transfer is investigated in the absence of a magnetic field. The RHTCs corresponding to different sets of ($\theta$, $f_g$) are given in Fig. 7(a). The results reveal a strictly monotonic decreasing of RHTCs as $\theta$ is increased from 0° to 90°, which is mainly attributed to the symmetry breaking caused by the twisting effect of gratings. The similar attenuation of heat transfer caused by twist has been observed in other NFRHT systems [64-66, 68, 69], and they are all explained by the breaking of symmetry between two anisotropic surfaces. For larger $f_g$, the modulation effect of $\theta$ decreases and reduces to zero when the gratings convert into sheets ($f_g=1$). When $f_g$ is less than 1, the reflection characteristics of the two gratings are anisotropic. Hence, the heat transfer is greatly dependent on the matching degree of the two interfaces, which is dependent on the twisted angle $\theta$. As a result, twist action can largely reduce the heat transfer between the two graphene gratings. For large graphene filling factors, which means the grating configuration is mostly occupied by graphene strips, the anisotropy of graphene conductivity is so weak that the matching degree of the two interfaces is weakly influenced by the twisted angle. In Fig. 7(b), RHTC modulation factors of graphene filling factor $\eta$ ($f_g$)=($h(f_g,\theta)$-$h(1,\theta)$)/$h(1,\theta)$ are given for different twisted angles. When the two graphene gratings are parallel ($\theta=0°$), $\eta(f_g)$ ranges from -0.75 to 0. When the graphene gratings are twisted ($\theta\neq0°$), the modulation range of $\eta(f_g)$ increases as the twisted angle $\theta$ increases. At different $f_g$, the modulation ability of $f_g$ is always greater for larger $\theta$, which means the twist action can contribute to the modulation ability of $f_g$. However, the modulation speed of $f_g$ is lower at low $f_g$ (0.1–0.8) and higher at high $f_g$ (0.8–1) with increasing $\theta$. In the absence of a magnetic field, the modulation ability of $f_g$ is very weak, the $\eta(f_g)$ is nearly unchanged, especially at low $f_g$ (0.1–0.6).

Fig. 8

According to the conductivity for $B=0$ T in Fig. 4(c), the graphene SPPs modes are hyperbolic in the absence of a magnetic field when graphene is patterned into gratings. Spectral RHTC and anisotropic energy transmission coefficients for different $\theta$ with $f_g=0.5$ are given in Fig. 8 to reveal the underlying mechanism of the attenuation caused by twist. The spectral RHTC obviously reveals that the radiative heat transfer becomes weaker as the twisted angle increases. Although the peak of spectral RHTC slightly increases for large $\theta$ (75°, 90°), it is decreasing in most part of the spectrum and the total heat transfer is reducing. In Fig. 8(b), the red and green dashed lines denote the dispersion relations of the two graphene gratings. Clearly, they are in coincidence for $\theta=0°$, which means the reflection characteristics of the two gratings are in good agreement. The contours show that the anisotropic energy transmission coefficients reach the maximum where the two dispersion relation curves intersect. The mismatch between the two dispersion relation curves of the two gratings increases as the twisted angle increases. Hence, the anisotropic energy transmission coefficients gradually decay and then the heat transfer is undoubtedly weakened.

## C. COMBINED EFFECT OF MAGNETIC FIELD AND GRATING

Fig. 9

In this section, the effects of magnetic field and grating are investigated simultaneously to reveal their combined effect. Fig. 9 has the same form as Fig. 7, whereas the magnetic field is applied to the graphene gratings this time and $B=15$ T. Comparing the results of RHTCs in Figs. 7(a) and 9(a), magnetic field endows the grating action ($\theta$ and $f_g$) with a higher modulation ability to modulate the NFRHT. Moreover, a unique enhancement region caused by grating is observed and denoted with dashed-yellow line in the figure. The region indicates that the NFRHT between graphene gratings can be larger than that of graphene sheets ($f_g$=1) at different twisted angles as a result of magnetic field. Moreover, the enhancement region of $f_g$ shrinks with increasing twisted angle. To demonstrate the modulation effect of grating in detail, we plot the $\eta(f_g)$ curves in Fig. 9(b) for different twisted angles. Similar to the results in the absence of magnetic field, the twist action results in a higher modulation range of graphene filling factors. Compared with the results in Fig. 7(b), the twisted angle has a greater influence on $f_g$ to modulate the NFRHT when a magnetic field is applied. The modulation speed of $f_g$ is much larger than that in zero field, especially at low $f_g$ (0.1–0.6). The unique enhancement region in Fig. 9(a) is illustrated with the inset in Fig. 9(b) and the shrinking effect caused by twist can be clearly observed. As explained in Fig. 6, the unique enhancement is mainly attributed to the great increasing of the near-zero modes, which are caused by the combined effect of magnetic field and grating. When the two gratings are twisted, the mismatch between the reflection characteristics weakens the effect of the near-zero modes. However, the enhancement region exists for different twisted angles. It is worth mentioning that for two parallel graphene gratings, when the graphene strips occupy more than 60% of the grating configuration, the NFRHT is basically equal to or larger than that of two graphene sheets.

Fig. 10

In Fig. 10, the detailed results considering magnetic field and twist are given. For $f_g$=0.9, the RHTCs corresponding to different sets of ($B$, $\theta$) are given in Fig. 10(a-1). The magnetic field has great effect on NFRHT when the gratings are twisted and the RHTCs decay with increasing $B$. To demonstrate the effect of twist clearly, we plot the modulation factors of twisted angle $\eta(\theta)=(h(B,\theta)-h(B,0))/h(B,0)$ for different magnetic fields in Fig. 10(a-2). The results reveal an obvious difference in $\eta(\theta)$ curves between zero-field and non-zero-field, which is caused by their different modes. When graphene strips greatly occupy the grating configuration, $f_g$=0.9, the near-zero modes, high-frequency hyperbolic and elliptic modes all exist, which are well demonstrated in Fig. 6. Moreover, the near-zero modes and high-frequency elliptic modes dominate the NFRHT. However, only hyperbolic modes exist for graphene gratings in the absence of magnetic fields, hence the twist action has little modulation effect on NFRHT, which can be clearly found in Fig. 7(a) for large $f_g$. In the presence of magnetic fields, the $\eta(\theta)$ curves differ from each other at different $\theta$, and the modulation ability at different $\theta$ is greater for larger *B*, which means that the modulation ability of twist can be tuned by the magnetic field at different twisted angles. The modulation ability of twist to tune the NFRHT can be greatly enhanced by the magnetic field, and the modulation range of twist angle nearly doubles when the magnetic field *B*=15 T is applied. To explain the attenuation trend caused by twist in the presence of a magnetic field, we provide the anisotropic energy transmission coefficients in Figs. 10(b) and 10(c) for different $\theta$ with $f_g$=0.9 and *B*=15 T. The frequencies of the first and second columns are respectively 0.005 $eV/\hbar$ and 0.18 $eV/\hbar$, which are the frequencies of the two peaks for the near-zero modes and high-frequency elliptic modes. The red and green dashed curves denote the dispersion relations for the upper and bottom graphene gratings, respectively. The energy transmission coefficients achieve the maximum near the intersections of the two curves. As a result, for $\theta$=0°, energy transmission coefficients are the largest because the dispersion relations are a perfect match. When $\theta$ is larger than 0°, the mismatch between the two dispersion relations increases and the energy transmission coefficients decrease accordingly. In comparison with the results of zero-field, the greater modulation abilities of twist in the presence of magnetic field are mainly caused by the simultaneous attenuation of the near-zero modes, high-frequency hyperbolic modes, and elliptic modes. The above results clearly confirm that the twist action and magnetic fields can jointly contribute to tune the NFRHT in graphene grating system.

## IV. CONCLUSION

We have systematically studied the NFRHT between two twisted graphene gratings based on the quantum Hall regime of magneto-optical graphene. For large chemical potentials, the pattern transformation from sheets to gratings weakens the magneto-optical characteristics of graphene. As a result, the NFRHT between graphene gratings can be greatly influenced by magnetic field only with low electronic density, which reveals the quantum Hall regime. Three graphene SPPs modes are observed in the system: near-zero modes, high-frequency hyperbolic modes, and elliptic modes. However, only hyperbolic modes work in the graphene grating system without a magnetic field. It is worth mentioning that elliptic SPPs modes are observed in the graphene grating system for the first time, which is caused by the combined effect of magnetic field and grating.

Analysis of the effects of magnetic field, twisted angle, and graphene filling factor on the NFRHT reveals some unique phenomena. The following conclusions are obtained. (i) The magnetic field has great modulation effect on the NFRHT, no matter for parallel or twisted graphene gratings. By changing the strength of magnetic field, the positions and intensities of the modes can be modulated, and hence the NFRHT can be tuned accordingly. (ii) The near-zero modes can be greatly enhanced by the combined effect of grating and magnetic field, rendering magnetic graphene a promising candidate for the thermal communication at ultra-low frequency. Interestingly, the existence of the near-zero modes results in a unique enhancement of heat transfer for strong magnetic field and large graphene filling factors, no matter for any twisted angle. It is worth mentioning that the unique enhancement caused by modes at such ultra-low frequency has never been observed in graphene system before. (iii) The magnetic field endows the grating action (graphene filling factors and twisted angles) with a higher modulation ability to modulate the NFRHT compared with zero-field. Moreover, the modulation ability of twist can be tuned by the magnetic field at different twisted angles. The combined effect of magnetic field and grating provides graphene devices with a tunable way to realize energy modulation and multi-frequency thermal communications.

**Conflict of Interest**

The authors declare no competing financial interest.

**Acknowledgements**

The supports of this work by the National Natural Science Foundation of China (No. 51976044, 51806047) are gratefully acknowledged. The authors thank Oleg Kotov and L.X. Ge for fruitful discussions. M. A. acknowledges support from the Institute Universitaire de France, Paris, France (UE).

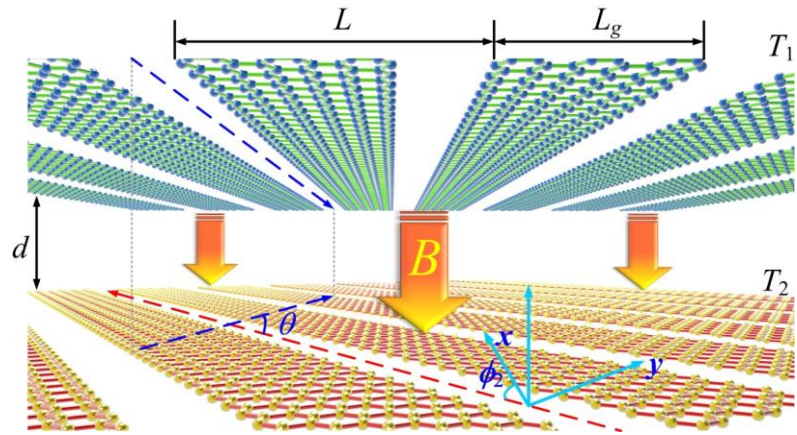

Fig. 1. Schematic of NFRHT between two twisted gaphene gratings composed of arrays of strips, whose periodicity and width are $L$ and $L_g$, respectively. The vacuum distance between them is denoted as $d$. The two sheets are kept at temperature $T_1$ and $T_2$. The main axes of the two gratings are labeled with the blue and red-dashed lines with twisted angle $\theta$. A magnetic field $B$ is applied perpendicularly to the gratings.

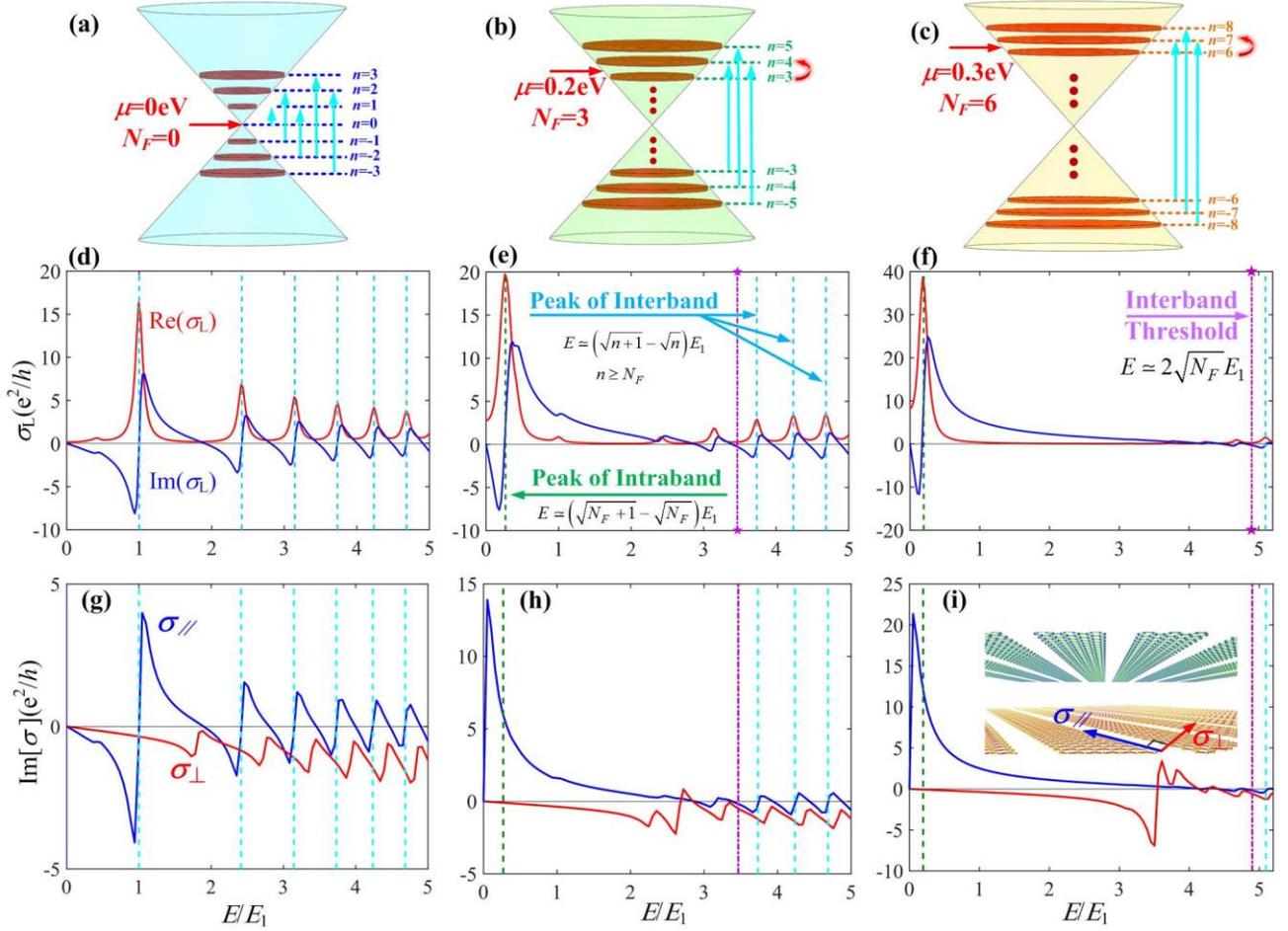

Fig. 2. Schematic of Dirac cones and different LLs in a magnetic field $B$=10T ($E_1 \approx 0.11$ eV) for (a) $\mu$=0 eV, (b) 0.2 eV and (c) 0.3 eV: The energies of the LLs are shown for different indices $n$ as red disks for both positive and negative Dirac cones, with the allowed interband transitions (vertical arrows in blue) and intraband transitions (crooked arrows in red). (d)-(f): The longitudinal conductivity of graphene $\sigma_L$ as a function of $E/E_1$, with $E = \hbar\omega$ and $E_1 = \sqrt{2}\hbar v_F / l_B$ for (d) $\mu$=0 eV, (e) 0.2 eV and (f) 0.3 eV; (g)-(i): Im[$\sigma_\parallel$] and Im[$\sigma_\perp$] with strip periodicity $L$=20 nm, filling factor $f_g$=0.5 for (g) $\mu$=0 eV, (h) 0.2 eV and (i) 0.3 eV.

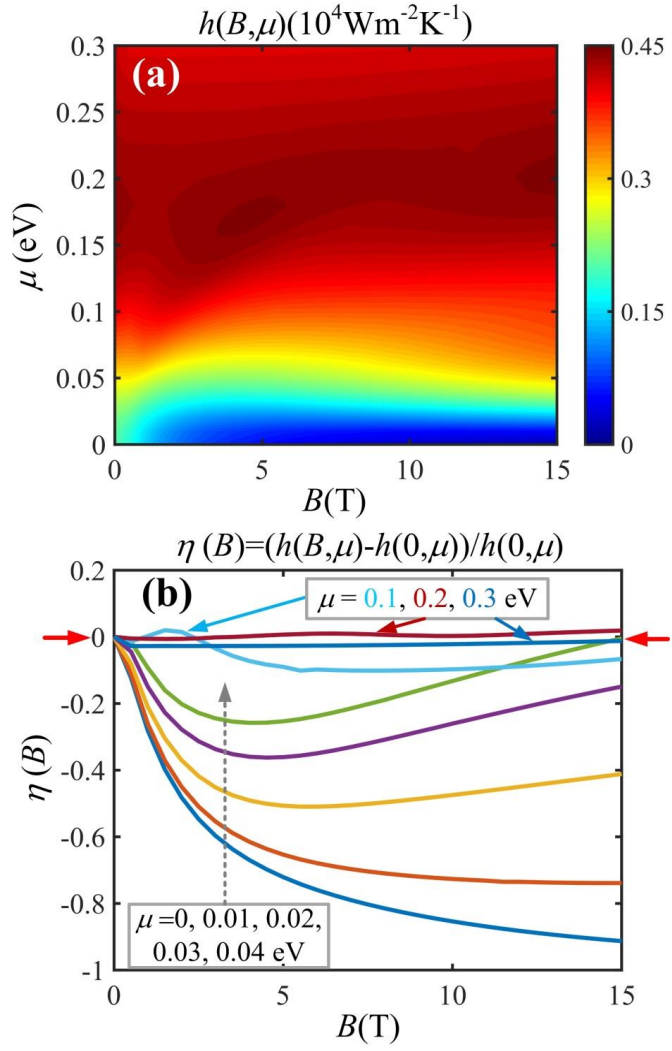

Fig. 3. (a) RHTCs corresponding to different sets of $(B,\mu)$ with $f_g$=0.5; (b) RHTC modulation factor of magnetic fields $\eta(B)=(h(B,\mu)-h(0,\mu))/h(0,\mu)$ for different $\mu$.

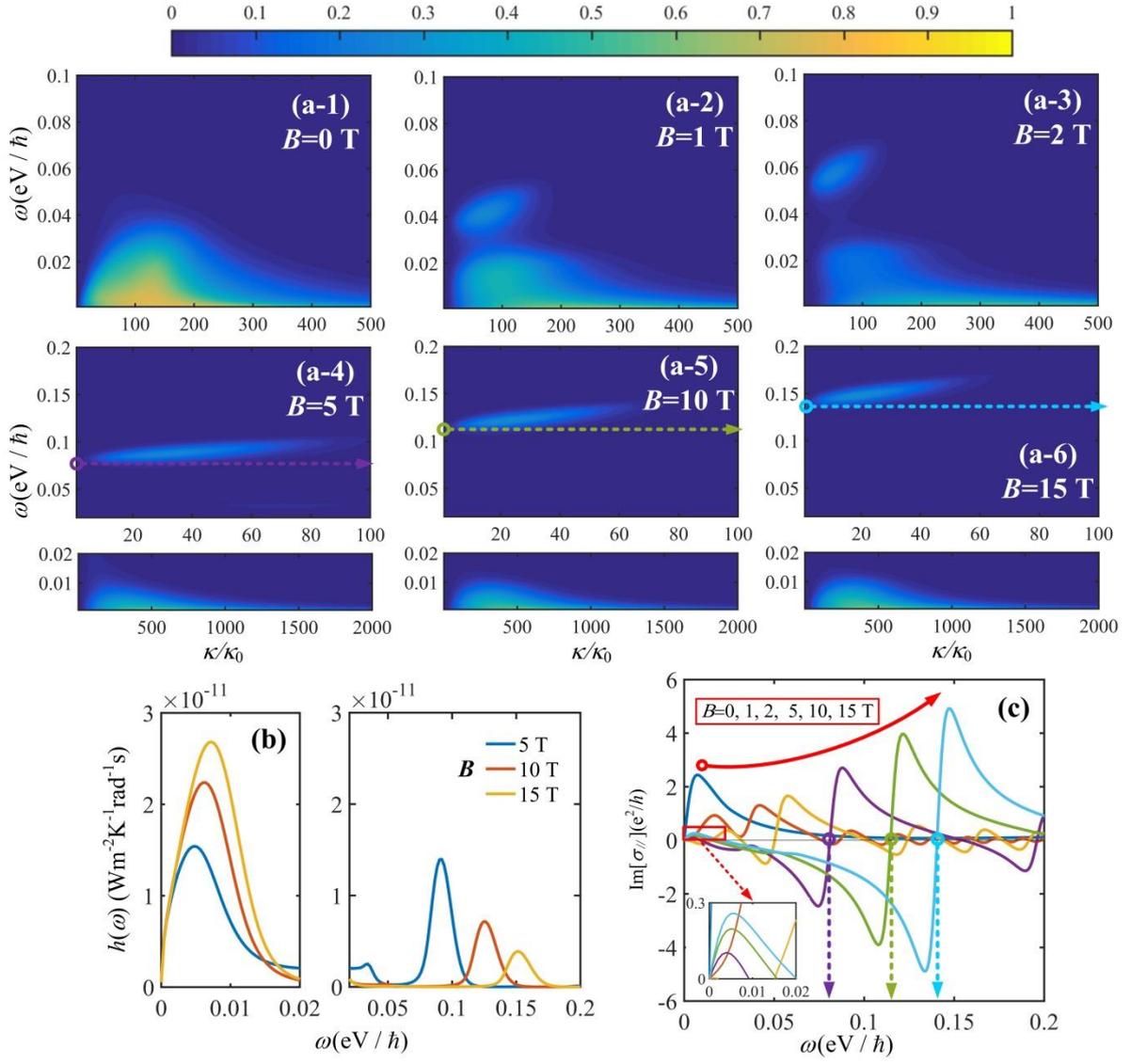

Fig. 4.   (a) Energy transmission coefficients, (b) spectral RHTCs and (c) Im[$\sigma_{//}$] for different $B$ with $f_g$=0.5

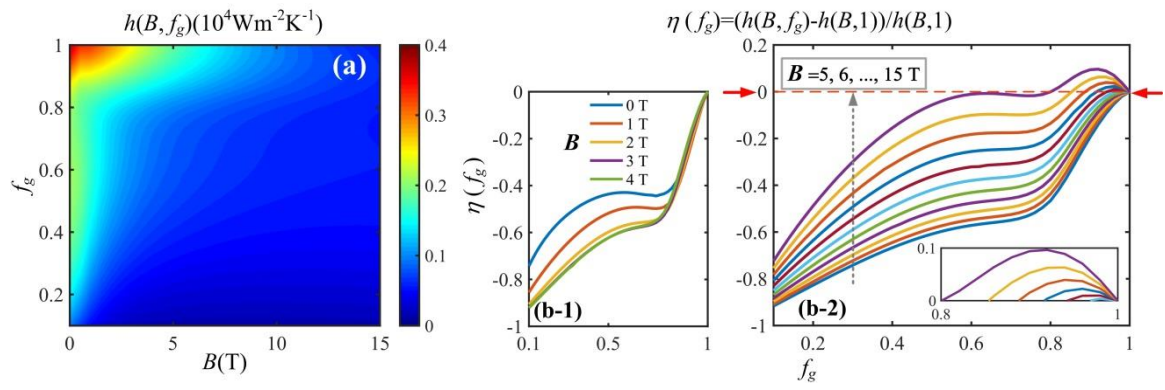

Fig. 5. (a): RHTCs corresponding to different sets of ($B, f_g$); (b): RHTC modulation factor of graphene filling factor $\eta(f_g) = (h(B, f_g) - h(B,1))/h(B,1)$ for different $B$

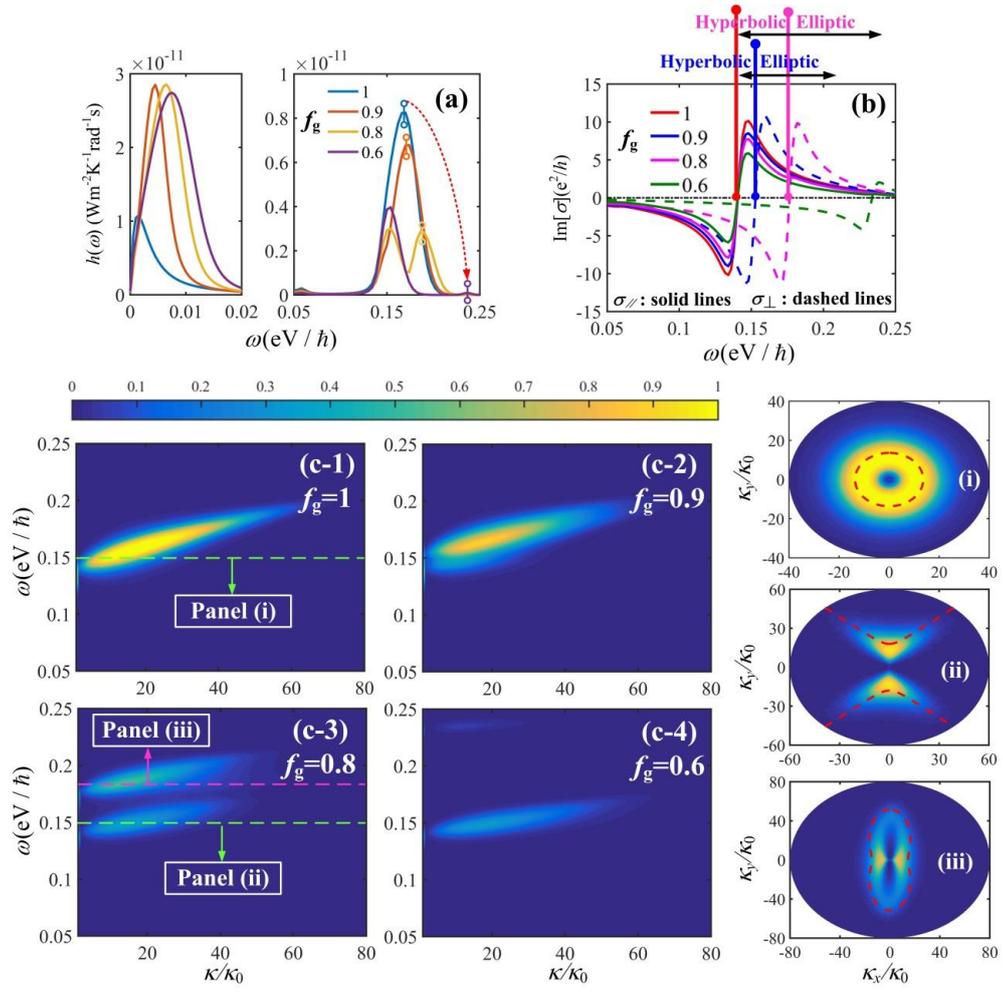

Fig. 6. (a) Spectral RHTCs, (b) Im[$\sigma$] and (c) energy transmission coefficients for different $f_g$ with $B$=15 T

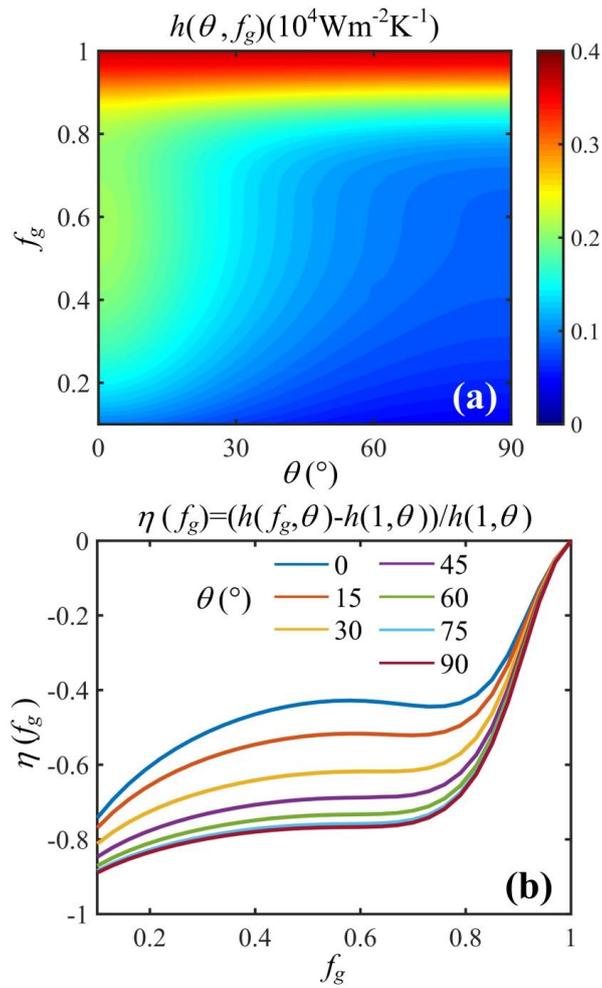

Fig. 7. In the absence of magnetic fields, (a) RHTCs corresponding to different sets of ($\theta$, $f_g$) and (b) RHTC modulation factor of graphene filling factor for different $\theta$

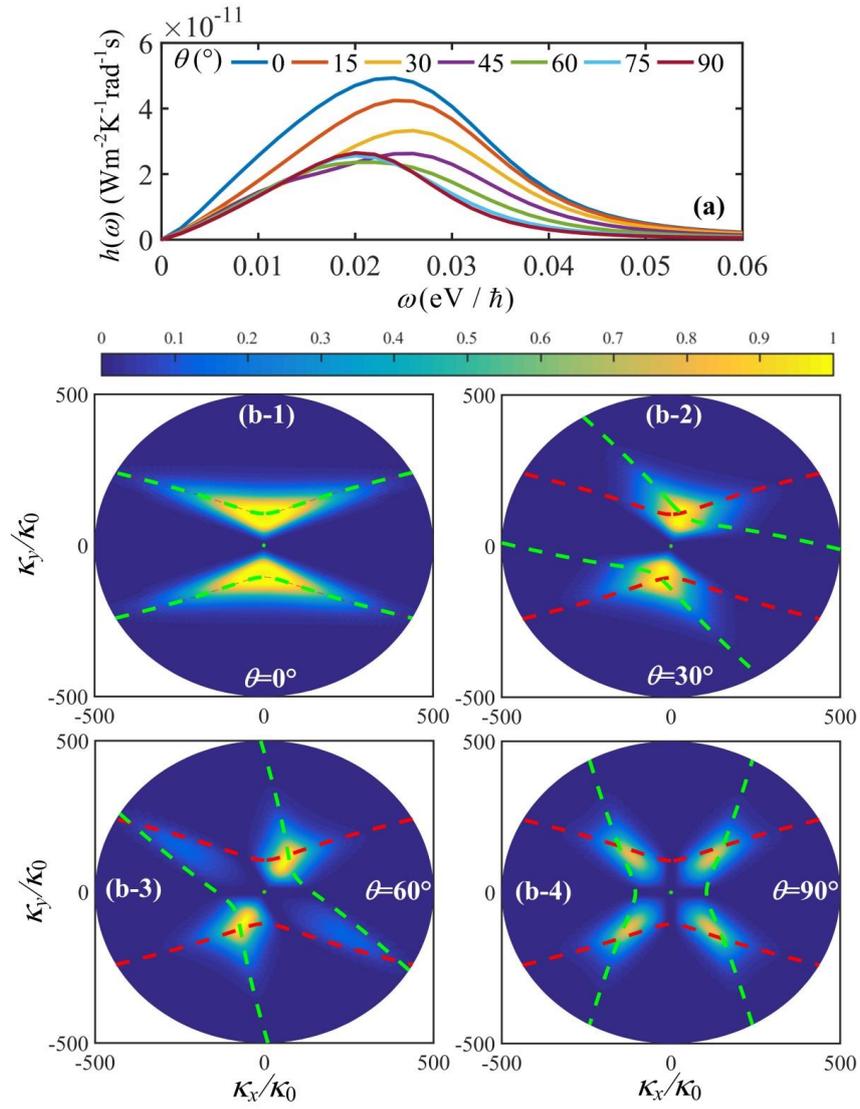

Fig. 8. (a) Spectral RHTCs and (b) anisotropic energy transmission coefficients for different twisted angles with $f_g$=0.5

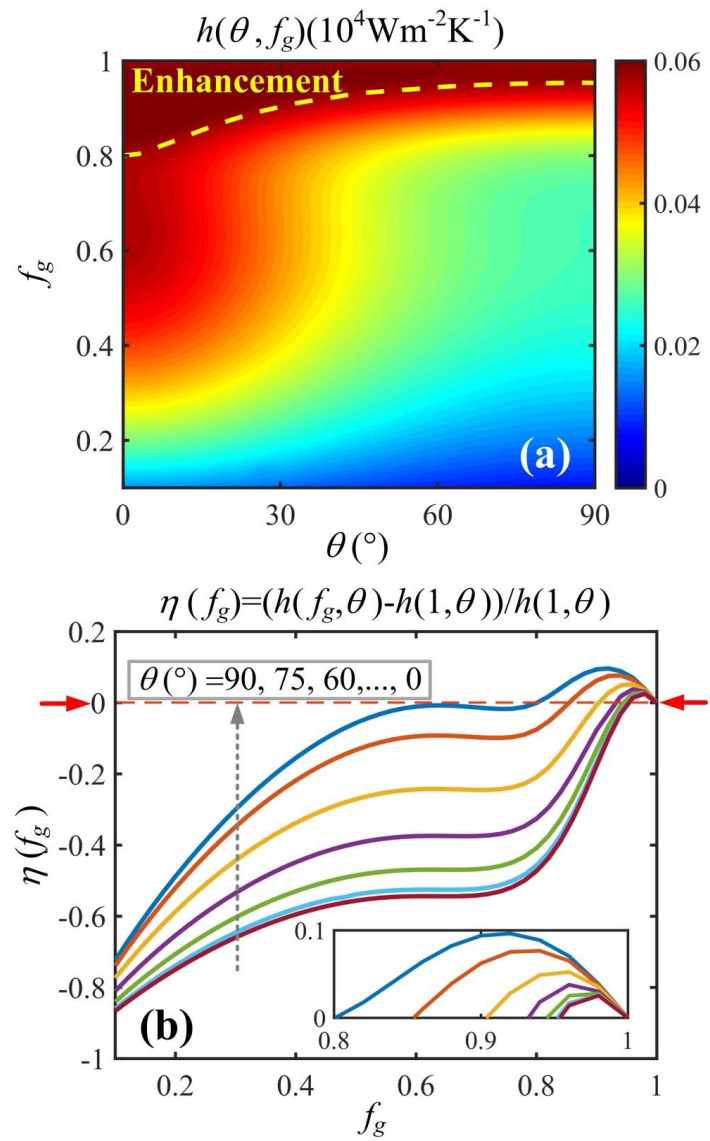

Fig. 9. In the presence of a magnetic field $B$=15 T, (a) RHTCs corresponding to different sets of ($\theta$, $f_g$) and (b) RHTC modulation factor of graphene filling factor for different $\theta$

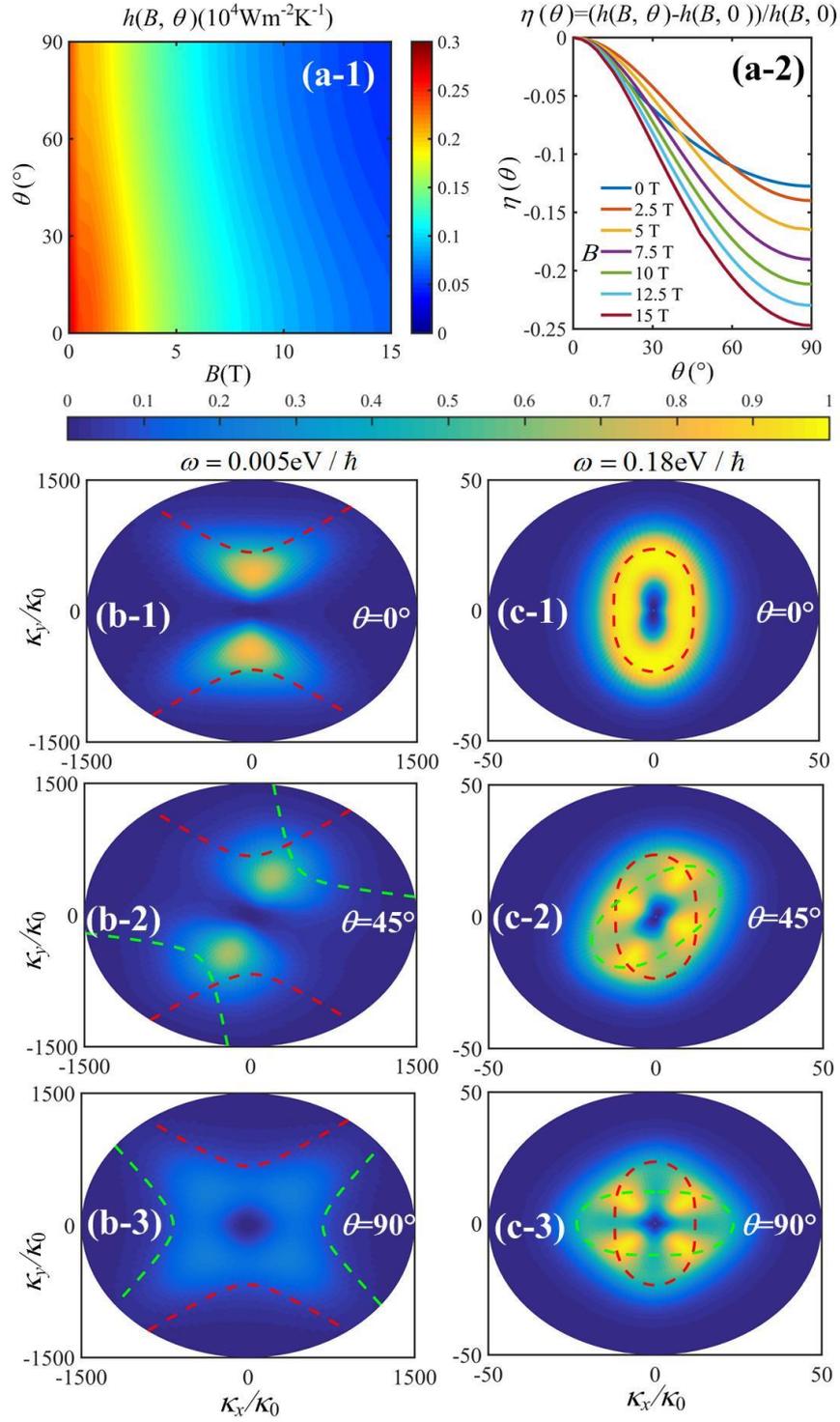

Fig. 10. For $f_g$=0.9, (a-1) RHTCs corresponding to different sets of ($B$, $\theta$) and (a-2) RHTC modulation factor of twisted angle $\theta$ for different $B$. (b)-(c) Anisotropic energy transmission coefficients for different $\theta$ with $B$=15 T.